# Ferrimagnet GdFeCo characterization for spin-orbitronics: large field-like and damping-like torques

*Héloïse Damas[1*], Alberto Anadon[1], David Céspedes-Berrocal[1,2], Junior Alegre-Saenz[1,2], Jean-Loïs Bello[1], Aldo Arriola-Córdova[1,2], Sylvie Migot[1], Jaafar Ghanbaja[1], Olivier Copie[1], Michel Hehn[1], Vincent Cros[3], Sébastien Petit-Watelot[1*] and Juan-Carlos Rojas-Sánchez[1*]*

[1]Université de Lorraine, CNRS, Institute Jean Lamour, F-54000 Nancy, France
[2]Universidad Nacional de Ingeniería, Rímac 15333, Peru
[3]Unité Mixte de Physique, CNRS, Thales, Université Paris-Saclay, 91767 Palaiseau, France

Corresponding authors: heloise.damas@univ-lorraine.fr, sebastien.petit@univ-lorraine.fr, Juan-Carlos.ROJAS-SANCHEZ@univ-lorraine.fr

H. Damas, Dr. A. Anadon, D. Céspedes-Berrocal, A.Y. Arriola-Córdova, J-L Bello, S. Migot, J. Ghanbaja, Dr. O. Copie, Prof. M. Hehn, Dr. S. Petit-Watelot, Dr. J.-C. Rojas-Sánchez.
Université de Lorraine, CNRS, Institute Jean Lamour, F-54000 Nancy, France
E-mail: heloise.damas@univ-lorraine.fr, sebastien.petit@univ-lorraine.fr, Juan-Carlos.ROJAS-SANCHEZ@univ-lorraine.fr

D. Céspedes-Berrocal, A.Y. Arriola-Córdova, J. Alegre-Saenz
Universidad Nacional de Ingeniería, Rímac 15333, Peru

Dr. V. Cros,
Unité Mixte de Physique, CNRS, Thales, Université Paris-Saclay, 91767 Palaiseau, France

Spintronics is showing promising results in the search for new materials and effects to reduce energy consumption in information technology. Among these materials, ferrimagnets are of special interest, since they can produce large spin currents that trigger the magnetization dynamics of adjacent layers or even their own magnetization. Here, we present a study of the generation of spin current by GdFeCo in a GdFeCo/Cu/NiFe trilayer where the FeCo sublattice magnetization is dominant at room temperature. Magnetic properties such as the saturation magnetization are deduced from magnetometry measurements while damping constant is estimated from spin-torque ferromagnetic resonance (ST-FMR). We show that the overall damping-like (DL) and field-like (FL) effective fields as well as the associated spin Hall angles can be reliably obtained by performing the dependence of ST-FMR by an added dc current. The sum of the spin Hall angles for both the spin Hall effect (SHE) and the spin anomalous Hall effect (SAHE) symmetries are: $\theta_{DL}^{SAHE} + \theta_{DL}^{SHE} = -0.15 \pm 0.05$ and $\theta_{FL}^{SAHE} + \theta_{FL}^{SHE} = 0.026 \pm 0.005$. From the symmetry of ST-FMR signals we find that $\theta_{DL}^{SHE}$ is positive and dominated by the negative $\theta_{DL}^{SAHE}$. The present study paves the way for tuning the different symmetries in spin conversion in highly efficient ferrimagnetic systems.

## 1. Introduction

In the last years, ferrimagnets have attracted growing interest for their potential utility in spintronic devices [1]. In particular, GdFeCo ferrimagnetic alloy is extensively studied as it exhibits a wide diversity of phenomena arising from the specific properties of rare earth-transition metal (RE-TM) ferrimagnets. Furthermore, the two antiferromagnetically coupled sublattices have a different response to external stimuli and the spin-orbit coupling (SOC) of the Gd 5d state allows the interplay between charge, spin, and orbital transport. The different relaxation times of these two coupled sublattices are thought to be responsible for the all-optical helicity-independent switching (AO-HIS) in GdFeCo demonstrated for almost a decade [2,3]. AO-HIS has also been recently observed in TbCo [4]. Nowadays, GdFeCo is used to perform the AO-HIS of Co/Pt [5–7] or CoNi/Pt [8] ferromagnetic multilayers. Moreover, it is possible to tune the Dzyaloshinskii-Moriya interaction in thin GdFeCo ferrimagnetic alloys [9], a relevant property for skyrmions formation. It has been shown that GdFeCo ferrimagnet also hosts large self-induced spin-orbit torque, or self-torque [10,11], with recent theoretical advances [12,13]. Ferro and ferrimagnetic materials are the source of spin currents with different symmetries [10,12] coming from the spin anomalous Hall effect (SAHE) [14–16] and the spin Hall effect (SHE) [17]. In the SAHE, the spin polarization of the spin current $J_s^{SAHE}$ is parallel to the magnetization while in the SHE it is perpendicular to both the injected charge current and the produced spin current $J_s^{SHE}$. A giant overall spin Hall angle for SAHE-like and SHE-like symmetries has been reported in a Gd-rich GdFeCo/Cu at room temperature [10]. Sizable interconversion efficiencies have also been reported for other magnetic materials such as NiFe [18–20] and CoFeB [15,21]. In the present work, we study room temperature FeCo-rich GdFeCo in a //$Gd_{25}Fe_{65.6}Co_{9.4}$(8 nm)/Cu(4 or 6 nm)/$Ni_{81}Fe_{19}$(4 nm) trilayer by structural, magnetic and spintronics characterization. We use two complementary ST-FMR techniques to reveal the signs and magnitudes of the contributions coming from the different spin current symmetries in GdFeCo. Namely, the modulation of the damping along with the shift of resonance field to extract the overall parameters (sum of the SHE-like and SAHE-like contributions) and the symmetry of the ST-FMR signal which is sensitive only to the SHE-like parameters. We found that damping-like (DL) SAHE spin Hall angle, $\theta_{DL}^{SAHE}$, is negative for FeCo-rich GdFeCo. In contrast, the DL SHE-like symmetry, is positive.

## 2. Structural and chemical characterization

Samples were grown on thermally oxidized Si wafers using dc magnetron sputtering at room temperature with an Ar gas pressure of 3 mTorr and base pressure of $1 \times 10^{-7}$ Torr. GdFeCo ($Gd_{25}Fe_{65.6}Co_{9.4}$) was co-deposited using separate Gd, Co, and Fe targets. All the samples in the present study were capped with 3 nm of naturally oxidized Al. The composition was controlled by varying the sputter gun power on each target. The deposition rate was calibrated by X-ray reflectivity and lift-off and profilometer measurements of the thickness. In order to perform structural characterization, a thin lamella was extracted by focused ion beam (FIB) milling using an FEI Helios Nanolab dual-beam 600i.

Transmission electron microscopy (TEM) investigations were carried out using a JEM - ARM 200F Cold FEG TEM/STEM (Scanning TEM) operating at 200 kV, coupled with a GIF Quantum 965 ER and equipped with a spherical aberration (Cs) probe and image correctors (point resolution 0.12 nm in TEM mode and 0.078 nm in Scanning TEM (STEM) mode). High-Resolution TEM (HRTEM) micrographs were performed to study the atomic structure of the deposit layers as shown in **Figure 1a**. The Fast Fourier Transformation (FFT) patterns in **Figure 1b,c** confirm that the Cu/NiFe layers are [111] textured along the growth direction while GdFeCo is amorphous as evidenced by the diffuse rings. Electron Energy Loss Spectroscopy (EELS) maps were carried out systematically on the different samples and confirm the nominal composition and thickness of the different materials (**Figure 1d**). We also evidence a slight GdFeCo composition variation along the growing direction as usually observed in RE-TM ferrimagnets [10,11,22]. The EELS maps displayed in **Figure 1d** were performed with 1ev/channel and a step of 0.3 nm.

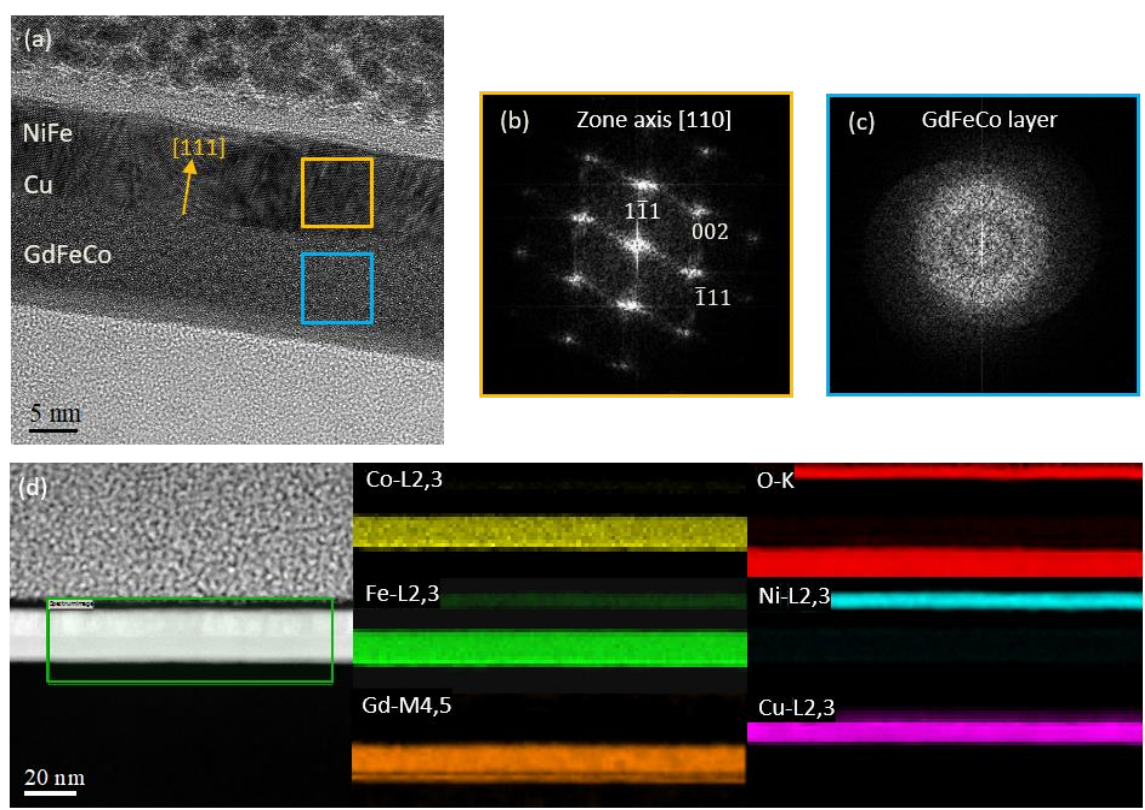

**Figure 1. TEM/STEM characterization of $Gd_{25}Fe_{65.6}Co_{9.4}(8)/Cu(6)/Ni_{81}Fe_{19}(4)/AlOx$.** (a) HRTEM micrograph of the deposited layers. The yellow (blue) square shows where the FFT analysis have been performed on Cu/NiFe (GdFeCo). The FFT patterns (b) and (c) indicate the [111] growth direction of textured Cu/NiFe and that GdFeCo is amorphous. (d) High Angle Annular Dark Field (HAADF)-STEM micrograph and the corresponding individual EELS elemental maps obtained from the green rectangle area in the HAADF micrograph. Co (yellow), Fe (green), Gd (orange), O (red), Ni (cyan), Cu (pink).

## 3. Magnetic characterization: magnetic anisotropies in GdFeCo

### 3. 1. SQUID magnetometry

Magnetization loops were performed at room temperature on a //GdFeCo(8)/Cu(6)/NiFe(4) stack with the applied field parallel and perpendicular to the field plane. Both $M(H)$ measurements are displayed in **Figure 2a** showing an open hysteresis loop. This confirms that the NiFe magnetization direction $\hat{m}_{\text{NiFe}}$ lies in the plane of the sample while that of GdFeCo, $\hat{m}_{\text{GdFeCo}}$, is spontaneously perpendicular to the film plane as shown in the inset. We assume that the 6 nm thick Cu layer decouples the two magnetic layers to extract their distinct saturation magnetization and saturation magnetic field. For NiFe, the saturation magnetization $M_s^{\text{NiFe}}$ is 625 kA/m and the saturation field $\mu_0 H_{\text{sat}-z}^{\text{NiFe}}$ to place $\hat{m}_{\text{NiFe}}$ out of the plane of the film is 0.85 T. In the case of GdFeCo, the saturation magnetization $M_s^{\text{GdFeCo}}$ is 115 kA/m and the saturation field $\mu_0 H_{\text{sat}-xy}^{\text{GdFeCo}}$ to align $\hat{m}_{\text{GdFeCo}}$ along the plane is about 0.13 T which are typical values for both NiFe and $Gd_{25}Fe_{65.6}Co_{9.4}$ at room temperature [23–25]. From the saturation field and magnetizations, we can also estimate the effective saturation magnetization for both magnetic materials, it results $M_{\text{eff}}^{\text{NiFe}} = 676$ kA/m and $M_{\text{eff}}^{\text{GdFeCo}} = 103$ kA/m. The relatively low perpendicular magnetic anisotropy of GdFeCo allows its magnetization to be easily placed along the plane of the film which is useful for ST-FMR measurements. The trilayer used in the next section has a 4 nm Cu spacer and displays a lower saturation field $\mu_0 H_{\text{sat}-xy}^{\text{GdFeCo}}$ to align $\hat{m}_{\text{GdFeCo}}$ along the plane, which is about ~0.047 T.

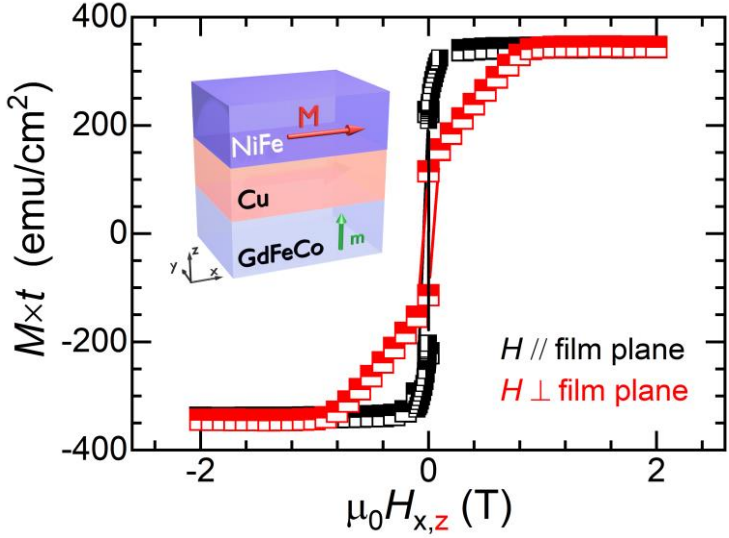

**Figure 2. Bulk magnetization data using a SQUID magnetometer.** (a) Magnetic hysteresis loop of the $Gd_{25}Fe_{65.6}Co_{9.4}(8)/Cu(6)/Ni_{81}Fe_{19}(4)$ trilayer. Magnetization values are normalized by the surface sample. To identify the different saturation fields and effective saturation magnetization, we consider that the magnetic layers are decoupled by the 6 nm of Cu. The inset shows a schematic of the sample with the spontaneous magnetization alignment of GdFeCo (out-of-plane) and NiFe (in-plane) according to SQUID results.

### 3. 2. Spin-torque FMR study

We perform ST-FMR measurements [26–31] on a GdFeCo(8)/Cu(4)/NiFe(4) trilayer to extract properties such as the damping constant $\alpha$ and the Landé g-factor of the magnetic layers. From Magneto optic Kerr effect measurements we have verified that this GdFeCo(8) is FeCo-rich at room temperature. The experimental setup is described in **Figure 3a**. A radiofrequency (rf) charge current, $i_{rf}$, is applied along the $\hat{x}$ direction and generates an oscillating Oersted field which triggers the magnetization precession at the resonance condition. A sweeping dc magnetic field $H_{dc}$ is applied in the *xy* plane of the device, at an angle of $\varphi_H$ with respect to the current line. At the resonance field $H_{res}$, a dc voltage $V_{mix}$ composed of a mixing of a symmetric and antisymmetric Lorentzians of amplitude $V_{sym}$ and $V_{anti}$ respectively can be measured using a bias tee. The measured mixed voltage displayed in **Figure 3b** can be fitted with the following general expression:

$$V_{\text{mix}} = V_{\text{offset}} + V_{\text{sym}} \frac{\Delta H^2}{\Delta H^2 + (H - H_{res})^2} + V_{\text{anti}} \frac{(H - H_{res})\Delta H}{\Delta H^2 + (H - H_{res})^2}, \quad (1)$$

where we consider an additional offset $V_{\text{offset}}$ and where $\Delta H$ is the linewidth. In **Figure 3b,** we observe the two resonance lines corresponding to the NiFe resonance (lower resonance field) and the GdFeCo resonance (higher resonance field). For the sake of clarity, it is only shown at 8, 12 and 14 GHz. Then, from broadband frequency dependence ST-FMR we can extract the effective saturation magnetization $M_{\text{eff}}$ (it results negative for systems where perpendicular magnetic anisotropy dominates over shape anisotropy), and the Landé g-factor considering the following expression:

$$f = \frac{\gamma}{2\pi}\sqrt{(H + H_{\text{uni}})(M_{\text{eff}} + H + H_{\text{uni}})}, \qquad (2)$$

where $\gamma = \frac{g\mu_B}{\hbar}$ is the gyromagnetic ratio and where $H_{\text{uni}}$ stands for a small in-plane uniaxial magnetic anisotropy. **Equation 2** applies for a thin film ferromagnetic layer with a magnetic field applied in the plane. We fix the NiFe Landé g-factor to 2.10. We determine the effective saturation magnetization of NiFe, $M_{\text{eff}}^{\text{NiFe}} = 569 \pm 1$ kA/m. The difference with previous SQUID results comes from the difference in Cu thickness which affects the NiFe anisotropy. We also evaluate a rather small $H_{\text{uni}} = -7 \pm 1$ Oe. We exploit the same **Equation 2** for GdFeCo resonance condition to determine the GdFeCo Landé g-factor and its effective saturation magnetization $M_{\text{eff}}^{\text{GdFeCo}}$. We obtain g = $2.87 \pm 0.04$, and $M_{\text{eff}}^{\text{GdFeCo}} = -37 \pm 4$ kA/m ($-46.5$ mT). The fitted experimental data is shown in **Figure 3c**. Finally, from the frequency dependence of the linewidth $\Delta H$, we calculate the Gilbert-type magnetic damping constant $\alpha$:

$$\Delta H = \Delta H_0 + \frac{2\pi f}{\gamma}\alpha, \qquad (3)$$

where $\Delta H_0$ is the $f$-independent contribution due to inhomogeneity. We have fixed g-Landé factor for both NiFe and GdFeCo. The fits on the measurements are shown in **Figure 3d**. The damping of in-plane NiFe is estimated as $\alpha^{NiFe} = 0.012 \pm 0.001$ which is about 8 times smaller than the damping of our out-of-plane $Gd_{25}Fe_{65.6}Co_{9.4}$ $\alpha^{GdFeCo} = 0.085 \pm 0.006$ but comparable with in-plane $Gd_{12.5}Fe_{76.1}Co_{11.4}$ [32]. Recently, it has been pointed out that actual or intrinsic damping in ferrimagnets is lower than that measured directly due to different spin density for each magnetic sublattice in GdFeCo and determined by domain wall mobility [33]. In the next section, we show how we can estimate the effective fields that drive the spin-orbit torque from $Gd_{25}Fe_{65.6}Co_{9.4}(8)/Cu(4)$ to $Ni_{81}Fe_{19}(4)$.

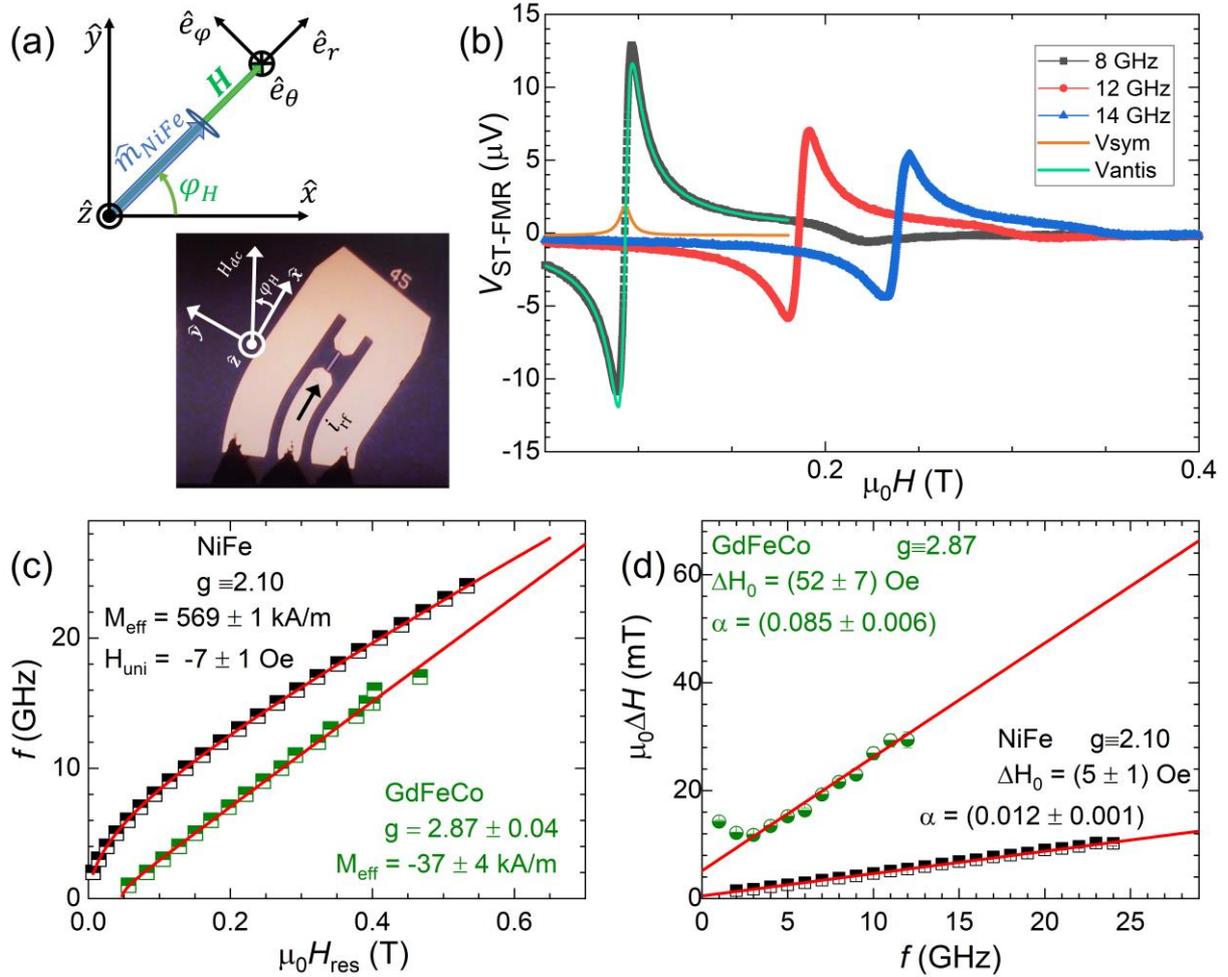

**Figure 3. Determination of $M_{eff}$ and damping using broadband ST-FMR in $Gd_{25}Fe_{65.6}Co_{9.4}(8)/Cu(4)/Ni_{81}Fe_{19}(4)$, and g-Landé factor for GdFeCo.** (a) Illustration of a typical ST-FMR device along with the dc magnetic field applied at $\varphi_H$ to the trilayer slab which is along $\hat{x}$. (b) Typical ST-FMR spectra at 8, 12 and 14 GHz. At a higher fields, the GdFeCo resonance line is observed. The symmetrical (orange) and antisymmetrical (green) voltage contributions are shown for NiFe at 8 GHz. Broadband frequency dependence of $H_{res}$ (c) and linewidth $\Delta H$ (d) are used to determine $M_{eff}$ and $\alpha$, respectively. Equation 2 is used for NiFe and GdFeCo layer in (c). For $Gd_{25}Fe_{65.6}Co_{9.4}$ (8 nm), we estimate $g = 2.87 \pm 0.04$ and $\alpha = 0.085 \pm 0.006$. Black (green) experimental data are obtained for the resonance of NiFe (GdFeCo) as depicted in (b).

## 4. Damping-like and field-like efficiencies determination by ST-FMR techniques

### 4. 1. Spin torque symmetries and ST-FMR signal

As discussed, there are two symmetries for spin current generation in magnetic materials, SAHE-like and SHE-like. When these spin currents are absorbed by another magnetic layer, they contribute to the total torque on the magnetization:

$$\mathbf{\Gamma}_{\text{tot}} = \mathbf{\Gamma}_{\text{SHE}} + \mathbf{\Gamma}_{\text{SAHE}}. \tag{4}$$

In the geometry of our ST-FMR measurements, the GdFeCo and NiFe magnetizations are both aligned with an angle of $\varphi_H$ with respect to the $\hat{x}$ axis. The spin polarization corresponding to the SHE-like symmetry, $\hat{\sigma}_{\text{SHE}}$, lies along the $\hat{y}$ axis regardless of the direction of both magnetizations. The spin polarization direction related to the SAHE-like spin current, $\hat{\sigma}_{\text{SAHE}}$, lies along the direction of $\hat{m}_{\text{GdFeCo}}$ which in turn is aligned with the equilibrium direction of the NiFe magnetization, $\hat{m}_{\text{NiFe}}$. The different contributions to the total torque can further be divided into two contributions, coming from the damping-like ($h_{\text{DL}}$) and the field-like ($h_{\text{FL}}$) effective fields:

$$\frac{\mathbf{\Gamma}_{\text{SHE}}}{\gamma M_S^{\text{NiFe}}} = h_{\text{DL}}^{\text{SHE}} \hat{m}_{\text{NiFe}} \times \left( \underbrace{\hat{\sigma}_{\text{SHE}}}_{\hat{y}} \times \hat{m}_{\text{NiFe}} \right) + h_{\text{FL}}^{\text{SHE}} \hat{m}_{\text{NiFe}} \times \underbrace{\hat{\sigma}_{\text{SHE}}}_{\hat{y}}, \tag{5a}$$

$$\frac{\mathbf{\Gamma}_{\text{SAHE}}}{\gamma M_S^{\text{NiFe}}} = h_{\text{DL}}^{\text{SAHE}} \hat{m}_{\text{NiFe}} \times \left( \underbrace{\hat{\sigma}_{\text{SAHE}}}_{\hat{m}_{\text{GdFeCo}}} \times \hat{m}_{\text{NiFe}} \right) + h_{\text{FL}}^{\text{SAHE}} \hat{m}_{\text{NiFe}} \times \underbrace{\hat{\sigma}_{\text{SAHE}}}_{\hat{m}_{\text{GdFeCo}}}. \tag{5b}$$

The efficiency of the charge-to-spin current conversion is described by the spin Hall angles $\theta_{\text{DL(FL)}}^{\text{SAHE}}$ and $\theta_{\text{DL(FL)}}^{\text{SHE}}$. They are related to the SAHE and SHE-like effective fields generated by GdFeCo and acting on NiFe layer as follows [14,15,34,35]:

$$h_{\text{DL(FL)}}^{\text{SHE}} = \frac{\hbar}{2|e|} \frac{j_c^{\text{GdFeCo}}}{\mu_0 M_S^{\text{NiFe}} t_{\text{NiFe}}} \theta_{\text{DL(FL)}}^{\text{SHE}}, \tag{6a}$$

$$h_{\text{DL(FL)}}^{\text{SAHE}} = \frac{\hbar}{2|e|} \frac{(\hat{m}_{\text{GdFeCo}} \times \mathbf{J}_c^{\text{GdFeCo}}) \cdot \hat{z}}{\mu_0 M_S^{\text{NiFe}} t_{\text{NiFe}}} \theta_{\text{DL(FL)}}^{\text{SAHE}}, \tag{6b}$$

with $(\hat{m}_{\text{GdFeCo}} \times \mathbf{J}_c^{\text{GdFeCo}}) \cdot \hat{z} = \sin(\varphi_H) J_c^{\text{GdFeCo}}$ in our geometry, as depicted in **Figure 3a**.

We show in the following subsections that ST-FMR techniques can be useful tools to further study the sign and quantification of the different contributions. Indeed, the analytical expression for the longitudinal voltage obtained by ST-FMR measurements reads [15,27,36,37]:

$$V_{dc} = -\frac{\Delta R_{AMR}^{NiFe}}{2} \sin(2\varphi_H) I_{rf} \left( \chi'_{\varphi\theta} \delta h_\theta + \chi'_{\varphi\varphi} \delta h_\varphi \right), \tag{7}$$

where $\Delta R_{AMR}^{NiFe}$ is the anisotropic magnetoresistance amplitude, $\chi'_{\varphi\theta}$ and $\chi'_{\varphi\varphi}$ are respectively the real part of the $\varphi\theta$ and the $\varphi\varphi$ components of the susceptibility matrix of NiFe. And, $\delta h_\theta$ and $\delta h_\varphi$ are respectively the polar and azimuthal component of the exciting field $\delta h$ (whose expression is discussed in the next subsection). We can see that only the transverse components of the excitation fields contribute to the ST-FMR voltage. We will discuss the different contributions to the total torque: i) first considering the symmetries of **Equation 7**, and ii) adding a dc current which will modify the susceptibility components.

We highlight here that all the equations and signs that our model describes have been verified by considering the results obtained in a //Pt(5)/NiFe(4) reference system. Namely, in this system, $\theta_{DL} = \theta_{DL}^{SHE} > 0$ and $\theta_{FL} = \theta_{FL}^{SHE} > 0$ (with a negative Oersted field).

### 4. 2. Symmetry of the ST-FMR signal

The NiFe magnetization resonance is triggered by the rf current induced Oersted field, and the spin torques described in **Equation 5a** and **Equation 5b**. We gather the different contributions under the general term of the exciting field, $\delta h$. Here, the delta means that the excitation is weak. The dynamics around the equilibrium position, which takes place in the $(\hat{e}_\theta, \hat{e}_\varphi)$ plane in spherical coordinates, is only sensitive to the polar and azimuthal components of the exciting field $\delta h_\theta$ and $\delta h_\varphi$. Since $\hat{\sigma}_{SAHE}$ lies along the NiFe magnetization equilibrium position $\hat{m}_{NiFe} = \hat{e}_r$, the associated SAHE effective fields do not contribute to the magnetization dynamics. On the contrary, the effective fields associated to the SHE-like symmetry contribute to the dynamics since $\hat{\sigma}_{SHE} \parallel \hat{y}$ and $\delta h_\theta = h_{DL}^{SHE}\cos(\varphi_H)$ and $\delta h_\varphi = \cos(\varphi_H)(h_{Oe} - h_{FL}^{SHE})$ [36,37]. Since at the resonance $\chi'_{\varphi\varphi}$ is an antisymmetric function of the applied magnetic field and $\chi'_{\varphi\theta}$ is a symmetric function, we can express the symmetrical voltage $V_{sym}$ amplitude and the antisymmetrical amplitude $V_{anti}$ introduced in **Equation 1** by replacing the suitable expressions in **Equation 7**:

$$V_{\text{sym}} = -\sin(\varphi_H) \frac{1}{4} \frac{I_{rf} \Delta R_{AMR}^{NiFe}}{\mu_0(2H + M_{eff}^{NiFe})} \frac{2\pi f}{\gamma} \frac{h_{DL}^{SHE}}{\Delta H}, \tag{8}$$

$$V_{\text{anti}} = -\sin(\varphi_H) \frac{1}{4} \frac{I_{rf} \Delta R_{AMR}^{NiFe}}{\mu_0(2H + M_{eff}^{NiFe})} \frac{2\pi f}{\gamma} \left[1 + \frac{M_{eff}}{H_{res}}\right]^{\frac{1}{2}} \frac{h_{Oe} - h_{FL}^{SHE}}{\Delta H}. \tag{9}$$

$V_{sym}$ ($V_{anti}$) only depends on $h_{DL}^{SHE}$ ($h_{Oe} - h_{FL}^{SHE}$) but the extraction of the effective fields using **Equation 8** and **Equation 9** is not trivial since the rf current has to be evaluated. Nevertheless, we can discuss the signs of the SHE effective fields. As depicted in **Figure 3b**, $V_{sym}$ is positive which means that $h_{DL}^{SHE} > 0$. $V_{anti}$ is negative, and thus $h_{FL}^{SHE} > 0$ assuming that the Oersted field is lower than the FL effective field.

### 4. 3. Adding a dc bias in ST-FMR: damping modulation and shift of $H_{res}$

When adding a dc bias to the previous ST-FMR measurement, a constant torque is applied on the oscillating magnetization which results in a change in the expression of its dynamical susceptibility matrix. This change induces a modulation of the linewidth and a shift in the resonant field, which can be both probed by the ST-FMR technique with an added dc bias. Because the susceptibility is related to the effective field along which the magnetization lies, only the spin polarizations with a projection along this effective field induce a change in the susceptibility. The modulation of damping technique is thus sensitive to both the SHE and SAHE-like symmetries and allows to extract overall parameters.

In the limit of low current densities where we can neglect strong heating contribution that deformed the linear behavior, we can modify the expressions developed for magnetic tunnel junctions [38,39], to apply it in our system [10,15,26]. For the modulation of the NiFe linewidth, it reads:

$$\frac{\partial \Delta H_{\text{NiFe}}}{\partial i_{dc}} = -\frac{f}{\gamma_{\text{NiFe}}} \frac{2}{(2H_{res}^{NiFe} + M_{eff}^{NiFe})} \frac{S_{\text{GdFeCo}}}{Wt_{\text{GdFeCo}}} \left( \underbrace{\widehat{\sigma}_{\text{SAHE}} \cdot \widehat{m}_{\text{NiFe}}}_{1} \frac{\partial h_{DL}^{SAHE}}{\partial J_c^{GdFeCo}} + \underbrace{\widehat{\sigma}_{\text{SHE}} \cdot \widehat{m}_{\text{NiFe}}}_{\sin(\varphi_H)} \frac{\partial h_{DL}^{SHE}}{\partial J_c^{GdFeCo}} \right), \quad (10)$$

where the left-hand term in the equation is the slope of the modulation of NiFe linewidth, $\gamma_{\text{NiFe}} = \frac{g_{NiFe}\mu_B}{\hbar}$. $S_{\text{GdFeCo}}$ accounts for the shunting of the GdFeCo layer by the other conductive layers, i.e., the current density flowing in GdFeCo layer is $J_c^{GdFeCo} = \frac{S_{\text{GdFeCo}}}{Wt_{\text{GdFeCo}}} i_{dc}$ with $W$ the width of the slab (10 μm). For simplicity, **Equation 10** can also be written in terms of the Hall angles using **Equation 6a,b** in the following way:

$$\frac{\partial \Delta H_{\text{NiFe}}}{\partial i_{dc}} = -\frac{f}{\gamma_{\text{NiFe}}} \frac{2}{(2H_{res}^{NiFe} + M_{eff}^{NiFe})} \frac{S_{\text{GdFeCo}}}{Wt_{\text{GdFeCo}}} \frac{\hbar}{2|e|} \sin(\varphi_H) \frac{\theta_{DL}^{SAHE} + \theta_{DL}^{SHE}}{\mu_0 M_s^{NiFe} t_{\text{NiFe}}}, \quad (11)$$

The slopes $\frac{\partial \Delta H_{\text{NiFe}}}{\partial i_{dc}}$ that account for the linewidth modulation at 8 GHz are displayed in **Figure 4b** for $\varphi_H = 135°$ and $\varphi_H = -45°$. The resistivities were determined independently through the dependence of the GdFeCo and Cu thicknesses for the different layers obtaining $\rho_{Cu} = 15$ μΩcm, $\rho_{GdFeCo} = 175$ μΩcm, and $\rho_{NiFe} = 40$ μΩcm. It follows $S_{\text{GdFeCo}} = 0.11$. We note that the slopes obtained when $\varphi_H = 135°$, for all the different frequencies measured, are opposite than the ones measured for the //Pt/NiFe reference sample (not shown). It indicates that the DL overall spin Hall angle, $\theta_{DL}^{SAHE} + \theta_{DL}^{SHE}$, is negative and opposite to the one of Pt where only the SHE is present. From the average of positive and negative dc fields, or 135° and -45°, and for 8, 12 and 14 GHz, we evaluate the overall DL efficiency $\theta_{DL}^{SAHE} + \theta_{DL}^{SHE} = -0.15 \pm 0.05$ for the FeCo-rich GdFeCo interfaced with Cu.

Furthermore, the same experiment also allows us to obtain the corresponding field-like values, $h_{FL}$ and $\theta_{FL}$. Based on the work of ref. [38,39], we also obtain the following expression that accounts for the linear displacement of the resonance field with an added dc current:

$$\frac{\partial H_{res}^{NiFe}}{\partial i_{dc}} = \frac{S_{GdFeCo}}{W t_{GdFeCo}} \left( \underbrace{\hat{\sigma}_{SAHE} \cdot \hat{m}_{NiFe}}_{1} \frac{\partial h_{FL}^{SAHE}}{\partial J_c^{GdFeCo}} + \underbrace{\hat{\sigma}_{SHE} \cdot \hat{m}_{NiFe}}_{\sin(\varphi_H)} \frac{\partial h_{FL}^{SHE}}{\partial J_c^{GdFeCo}} \right) - \underbrace{\hat{\sigma}_{SHE} \cdot \hat{m}_{NiFe}}_{\sin(\varphi_H)} \frac{\partial h_{Oe}}{\partial i_{dc}}, \quad (12)$$

where $h_{Oe}$ is the Oersted field which lies along the $-\hat{y}$ direction in the geometry of our system. Its amplitude can be approximated with $h_{Oe} = -\frac{1}{2}(j_c^{GdFeCo} t_{GdFeCo} + j_c^{Cu} t_{Cu})$. **Equation 12** reads in terms of the FL Hall angles (**Equation 6a,b**):

$$\frac{\partial H_{res}^{NiFe}}{\partial i_{dc}} = \sin(\varphi_H) \left[ \frac{S_{GdFeCo}}{W t_{GdFeCo}} \left( \frac{\hbar}{2|e|} \frac{\theta_{FL}^{SAHE} + \theta_{FL}^{SHE}}{\mu_0 M_s^{NiFe} t_{NiFe}} \right) - \frac{\partial h_{Oe}}{\partial i_{dc}} \right], \quad (13)$$

The slope obtained from the shift of the resonance field vs. $i_{dc}$ is displayed in **Figure 4c** for different frequencies. We observe that the slope is frequency-independent in agreement with **Equation 13**. Moreover, the slope has the same sign as the one in the //Pt/NiFe reference system. That implies that if there is any FL contribution on the GdFeCo/Cu/NiFe system studied here it has the same sign as for the //Pt/NiFe. The slope is evaluated as $\frac{\partial H_{res}^{NiFe}}{\partial i_{dc}} = 0.037$ T/A. The Oersted field is approximated as $\frac{\partial h_{Oe}}{\partial i_{dc}} = -0.0476$ T/A. Finally, considering **Equation 13**, the overall FL efficiency is assessed as $\theta_{FL}^{SAHE} + \theta_{FL}^{SHE} = 0.026 \pm 0.005$. This value has the same sign and is comparable to the one measured in NiFe/Pt [37,40]. We have independently measured a control //Cu/NiFe sample without a sizable effect. We can therefore exclude the Cu/NiFe interface as the origin behind the FL measured in GdFeCo/Cu/NiFe. The sizable overall FL value would indicate that even though GdFeCo is not in contact with NiFe, a significant FL contribution can still be detected. The origin of the FL effect in the trilayer is not clear at this stage.

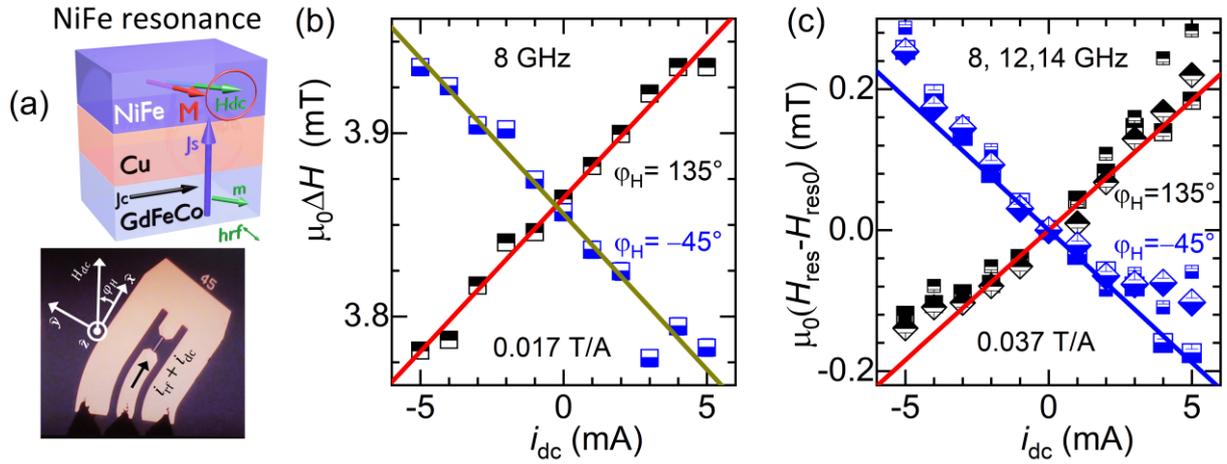

**Figure 4. Damping modulation and Resonance field shift**. (a) Schematic of the NiFe resonance condition with additional $i_{dc}$ current injected. (b) $i_{dc}$ dependence of the NiFe linewidth for a rf frequency of 8 GHz. (c) Resonance field shift vs. $i_{dc}$ for three frequencies. Unlike the damping or linewidth modulation, we can see that the resonance field shift is frequency independent.

## 5. Discussion and conclusions

The overall efficiencies for FeCo-rich GdFeCo/Cu/NiFe are evaluated $\theta_{DL}^{SAHE} + \theta_{DL}^{SHE} = -0.15 \pm 0.05$ and $\theta_{FL}^{SAHE} + \theta_{FL}^{SHE} = 0.026 \pm 0.005$. For sake of comparison, the SAHE efficiency of a ferromagnet such as CoFeB is $\theta_{SAHE}^{CoFeB} = -0.14$ [15], and the SHE efficiency of Pt heavy metal is $\theta_{SHE}^{Pt} = 0.056 - 0.076$ [29,41,42]. Seki *et al*. show in FePt that DL $\theta_{SAHE+SHE}^{FePt} = 0.25$ from the linewidth modulation [43].

The damping-like SAHE contribution dominates over the SHE one: $|\theta_{DL}^{SAHE}| > |\theta_{DL}^{SHE}|$ with a negative SAHE contribution for FeCo-rich GdFeCo, and a positive SHE contribution. We also show that the field-like SHE contribution is positive. However, we cannot estimate the individual value of each contribution. We perform the same experiments at 15 K where our ferrimagnet is Gd-rich and its magnetization aligns in-plane with a field above 0.4 T. From the sign of the symmetric contribution we confirm that SHE remains positive when crossing the magnetic compensation temperature. This is consistent with the fact that the SHE does not depend on the GdFeCo magnetic properties. In contrast, we cannot conclude of any $\theta_{DL}^{SAHE}$ sign change because the modulation of linewidth experiments at 15 K is hidden by others effect that are out of the scope of this study. However, the large variation in absolute value between these results and the one previously reported, for a Gd-rich GdFeCo at room temperature, $|\theta_{DL}^{SAHE} + \theta_{DL}^{SHE}| = 0.80 \pm 0.05$ [10], suggest that the sign of $\theta_{DL}^{SAHE}$ changes between FeCo-rich and Gd-rich samples. If so, the opposite DL-SAHE sign for FeCo-rich GdFeCo might indicate that the SAHE spin polarization comes always from the same magnetic sublattice. Despite that, further studies could be carried out to confirm that.

GdFeCo can thus generate efficient spin currents and the different symmetries allow this material to be used in a wide variety of devices for spintronics. For instance, the SHE spin current can generate self-torque [10] and can be used for the electrical switching of the magnetization, as shown in epitaxial FePt [44] or CoTb [45]. Also, the total spin current (SAHE+SHE) can be used to induce a torque on another magnetic layer or for the manipulation of skyrmions.

In summary, we have studied FeCo-rich GdFeCo/Cu/NiFe heterostructure at room temperature. First, structural, and chemical analyses were performed by HRTEM and EELS. Then, the magnetic properties and the relevant spin-orbitronics parameters were determined by combining magnetometry, spin-torque ferromagnetic resonance and additional dc current dependence. The overall damping-like and field-like efficiencies, which include the SHE-like and the SAHE-like symmetries, are $\theta_{DL}^{SAHE} + \theta_{DL}^{SHE} = -0.15 \pm 0.05$ and $\theta_{FL}^{SAHE} + \theta_{FL}^{SHE} = 0.026 \pm 0.005$ at room temperature. We show that SAHE dominates over SHE contribution on the DL torque. Furthermore, this study shows that the SHE contribution does not change sign when crossing the magnetic compensation temperature while SAHE may change sign depending on the dominant sublattice of the ferrimagnet. All this underlines the importance of GdFeCo, and RE-TM ferrimagnets in general, as promising materials in spintronics for the exploitation of their strong spin-orbit torque.

## Data availability

The data that support the findings of this study are available from the corresponding author on reasonable request.

## Acknowledgements

We acknowledge A. Fert for fruitful discussions. This work was supported from Agence Nationale de la Recherche (France) under contract ANR-19-CE24-0016-01 (TOPTRONIC), ANR-20-CE24-0023 (CONTRABASS), and ANR-17-CE24-0025 (TOPSKY), from the French PIA project "Lorraine Université d'Excellence", reference ANR-15IDEX-04-LUE and by the « SONOMA» project co-funded by FEDER-FSE Lorraine et Massif des Vosges 2014-2020, a European Union Program. DCB and JAS also thanks 2019 and 2021 Master-LUE program internship. Devices in the present study were patterned at MiNaLor clean-room platform which is partially supported by FEDER and Grand Est Region through the RaNGE project.